\documentclass[12pt,a4paper]{article}
\usepackage{graphicx}
\begin{document}

\pagestyle{empty}

\begin{flushright}
{\bf UA/NPPS-06-02}\\
{\bf McGILL-02-30}
\end{flushright}

\vglue 2cm
\begin{center} \begin{large} \begin{bf}
A MODEL FOR THE COLOR GLASS CONDENSATE VERSUS JET QUENCHING
\end{bf} \end{large} \end{center}
\vglue 0.35cm
{\begin{center}
A.P.\ Contogouris$^{(+) (*)}$,
F.K. Diakonos$^{(*)}$ and
P.K. Papachristou$^{(*)}$
\end{center}}
\parbox{6.4in}{\leftskip=1.0pc
{(+) \textit{Department of Physics, McGill University, Montreal, Quebec,
H3A 2T8, CANADA}}\\
\vglue -0.25cm
{(*) \textit{Nuclear and Particle Physics, University of Athens, Panepistimiopolis,
Athens 15771, GREECE}}}
\vglue 1cm
\begin{abstract}
A model for the Color Glass Condensate as opposed to jet quenching is proposed for the
explanation of the presently available RHIC data. Good fits to these data are presented. A
clear way to distinguish between the two possible explanations is also given.
\end{abstract}

\vglue 1cm \noindent 
{\large\bf 1.} Recent RHIC data on hadron ($\pi^0$) production at large transverse momentum $p_T$
in central $Au+Au$ collisions show a clear suppression of the rates \cite{adc02}. The usual explanation
is that the phenomenon is due to jet quenching, which thus makes a probe of gluon plasma \cite{bai97,gyu00}.
In the present work we propose an explanation of the same data as due to the Color Glass Condensate \cite{mcl02}.
Our account of the data provides also a way to distinguish between the two explanations.

At very high energies the number of partons (mainly gluons) in a nucleus grows very rapidly and eventually leads to
saturation \cite {gri83,mcl02}. We will attempt to express this saturation in the simplest way, by invoking expressions
used at small $x$.

\vglue 0.5cm \noindent
{\large\bf 2.} With $g(x,Q^2)$ the gluon distribution, at small $x$ ($P_{gg} (x) \to 2N_c /x$) a simple evolution equation is \cite{kwi95}:
\begin{equation}
\frac{{\partial g(x,Q^2 )}}{{\partial \ln Q^2 }} = \frac{{\alpha _s (Q^2 )}}{{2\pi }}\int\limits_x^1 {dy\frac{{2N_c }}{x}} g(y,Q^2 ) - \frac{{9\pi ^2 }}{{16R^2 }}\left( {\frac{{\alpha _s }}{{2\pi }}} \right)^2 \frac{6}{{Q^2 }}\int\limits_x^1 {dy\frac{{2N_c }}{{{x \mathord{\left/
 {\vphantom {x y}} \right.
 \kern-\nulldelimiterspace} y}}}} g^2 (y,Q^2 ).
\end{equation}
Here $R$ amounts to a free parameter, but will be taken as the radius of the quarks ($\simeq 0.1fm$). An approximate integration of the last term leads to the modified gluon distribution
\begin{equation}
\tilde F_{{g \mathord{\left/
 {\vphantom {g p}} \right.
 \kern-\nulldelimiterspace} p}} (x,Q^2 ) = F_{{g \mathord{\left/
 {\vphantom {g p}} \right.
 \kern-\nulldelimiterspace} p}} (x,Q^2 ) - \frac{{27\pi ^2 }}{8}\frac{{\alpha _s (Q^2 )}}{{2\pi }}\frac{1}{{R{}^2Q^2 }}G(x),
\end{equation}
where
\begin{equation}
G(x) = 2N_c \int\limits_x^1 {\frac{{dy}}{y}} F_{{g \mathord{\left/
 {\vphantom {g p}} \right.
 \kern-\nulldelimiterspace} p}}^2 (y,Q^2 ).
\end{equation}

\vglue 0.5cm \noindent
{\large\bf 3.}  The basic formula for $pp\to\pi^0+X$ is
\begin{equation}
E\frac{{d\sigma }}{{d^3 p}}(S,p_T ,\theta ) = \frac{{4K}}{{\pi x_T^2 }}\sum\limits_{a,b,c} {\int\limits_{x_1 }^1 {dx_a } \int_{x_2 }^1 {dx_b F_{{a \mathord{\left/
 {\vphantom {a p}} \right.
 \kern-\nulldelimiterspace} p}} (x_a )F_{{b \mathord{\left/
 {\vphantom {b p}} \right.
 \kern-\nulldelimiterspace} p}} (x_b )\frac{{d\hat \sigma }}{{dt}}D_{{{\pi ^0 } \mathord{\left/
 {\vphantom {{\pi ^0 } c}} \right.
 \kern-\nulldelimiterspace} c}} (z)\frac{\rho }{{\left( {1 + \rho } \right)^2 }}} } 
\end{equation}
where $K$ is a $K$-factor, here for simplicity taken K=2, $x_T=2p_T/\sqrt{S}$ and for $\theta=\pi/2$:
\begin{displaymath}
x_1  = \frac{{x_T }}{{2 - x_T }}~~~~x_2  = \frac{{x_T x_a }}{{2x_a  - x_T }}~~~~\rho  = \frac{{x_a }}{{x_b }} 
\end{displaymath}
and
\begin{equation}
z = x_T \frac{{1 + \rho }}{{2x_a }}. 
\end{equation}
Also
\begin{equation}
\frac{{d\hat \sigma }}{{dt}} = \frac{{\pi \alpha _s^2 (Q^2 )}}{{s^2 }}\Sigma (ab) 
\end{equation}
where e.g. $\Sigma (gg) = \frac{9}{2}\left( {3 - \frac{{tu}}{{s^2 }} - \frac{{us}}{{t^2 }} - \frac{{st}}{{u^2 }}} \right)$ etc.

\vglue 0.5cm \noindent
{\large\bf 4.} For $N_1N_2\to\pi^0+X$ one has
\begin{equation}
E\frac{{d\sigma _{N_1 N_2 } }}{{d^3 p}}(S,p_T ,\theta ) = \int {d^2 {\bf b}} \int {d^2 {\bf r}T_{N_1 } ({\bf b})T_{N_2 } ({\bf b} - {\bf r})E\frac{{d\sigma _{pp} }}{{d^3 p}}(S,p_T ,\theta )} 
\end{equation}
where $T_N({\bf b})$ is the Glauber thikness function ($=\int dz \rho_N({\bf r})$, $\rho_N=$density of nucleus $N=Au$) normalized as $\int d^2{\bf b}T_N({\bf b})=1$.
We use a gaussian $\rho _N ({\bf r}) \sim e^{ - r^2 /a^2 }$ and $b_{max}=4.7~{\rm fermi}$. The inclusive $E\frac{{d\sigma _{N_1 N_2 } }}{{d^3 p}}$ is augmented by an intrinsic transverse momentum of a gaussian with $\left\langle {k_T^2 } \right\rangle  = 1{\rm  GeV}^{\rm 2}$. For the parton distributions $
F_{{a \mathord{\left/
 {\vphantom {a p}} \right.
 \kern-\nulldelimiterspace} p}}$ we use the set CTEQ 5, leading order \cite{cteq5} and for the fragmentation functions we use the Binnewies et al., again leading order \cite{bin95}. Finally, we use $Q^2=p_T^2$ in Eqs (2) and (3).

\vglue 0.5 cm \noindent
{\large\bf 5.} Our results at 130 A $GeV$ for $Au+Au\to\pi^0$ production at $\theta = {{\pi} \over 2}$ are shown in Fig. 1 (solid line). The dashed line shows the results without the effects of the Color Glass Condensate. On the same figure we plot the results for jet quenching corresponding to opacity ${{L}\over {\lambda}}=3$ \cite{fai01} (dotted line). Both the solid and the dotted line account well for the data \cite{fai01}. However, at large $p_T$ ($p_T\geq 6 GeV$) the effect of the Color Glass Condensate tends to disappear and the solid line approaches the dashed line; this is due to the factor ${{1}\over{Q^2}}$ ($={{1}\over{p_T^2}}$), which appears in the modified gluon distribution. On the other hand, jet quenching remains below, and this gives the possibility to distinguish between the two mechanisms.

At very low $p_T$ ($<2 {\rm GeV}$) all lines diverge. Perturbative QCD is inapplicable and various effects, like recoil resummation, play a dominant role.

\vglue 0.5cm
\section*{Acknowledgments}
A number of helpful discussions with N. Antoniou, A. Bialas, S. Jeon and E. Mavrommatis, as well as an independent check of some of our results by Z. Merebashvili are gratefully acknowledged. The work was also supported by the Natural Sciences and
Engineering Research Council of Canada and by the Greek State Scholarships Foundation (IKY).

\newpage
\begin{center}
\includegraphics[height=22cm]{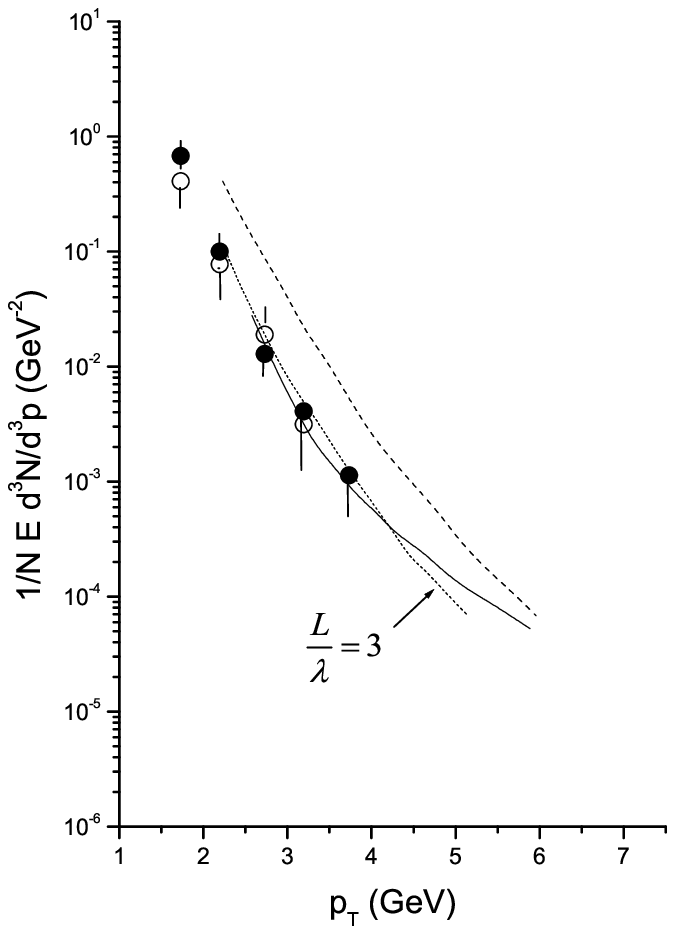}
\end{center}
\end{document}